\documentstyle[prl,aps,epsf,multicol]{revtex}
\begin{document}
%\RCS $Revision: 1.19 $
%\RCS $Date: 1998/01/04 12:24:41 $
\draft

\title{Frequency Dependent Conductance of a Tunnel Junction in the
Semiclassical Limit}
\author{Georg G\"oppert$^{1,2}$ and Hermann Grabert$^1$}
\address{$^1$Fakult\"at f\"ur Physik, Albert--Ludwigs--Universit{\"a}t, \\
Hermann--Herder--Stra{\ss}e~3, D--79104 Freiburg, Germany}
\address{$^2$Service de Physique de l'Etat Condens\'e, \\
           CEA-Saclay, 91191 Gif-sur-Yvette, France}

\date{\today}
\maketitle
\widetext

\begin{abstract}
The linear conductance of the a small metallic tunnel junction
embedded in an electromagnetic
environment of arbitrary impedance is
determined in the semiclassical limit. Electron tunneling is 
treated beyond the orthodox theory of Coulomb blockade phenomena by 
means of a nonperturbative path
integral approach. The frequency dependent conductance is 
obtained from Kubo's 
formula. The theoretical predictions are valid for high temperatures
and/or for large tunneling conductance and are found to explain 
recent experimental data.
\end{abstract}


%\raggedcolumns
%\begin{multicols}{2}
%\narrowtext 

\section{Introduction}
In recent years a great deal of experimental and theoretical work
\cite{Nato,Kastner} has
explored Coulomb charging effects in systems with
tunnel junctions. For
small tunneling conductance $G_T \ll G_K$, where $G_K=e^2/h$ is the
conductance quantum, the effects are theoretically well understood 
in terms of the ``orthodox'' perturbative approach in the tunneling
Hamiltonian \cite{Averin}. Present lithographic techniques make
the region of moderate to large 
tunneling 
conductance experimentally accessible \cite{JoyezSET,JoyezSJ,Toppari}.
In fact when using Coulomb blockade devices in 
thermometry \cite{Pekola} or
as highly sensitive electrometers \cite{Fulton},
a large tunneling conductance is often desirable in view of the larger
measuring signal. Several theoretical advances to describe strong
tunneling have been made recently. These more sophisticated theories 
roughly split 
into two groups. One may use higher order
perturbation theory in the tunneling Hamiltonian 
\cite{GrabertBOX,KoenigCOT,GeorgBOX}
and, based on this, a perturbative renormalization
group approach \cite{KoenigRG},
while the other set of papers starts from a formally exact path integral
expression \cite{SchoenREP} 
serving as a basis for 
analytical calculations \cite{ZaikinSET,WangBOX,GeorgSJ,GeorgSET},
Monte Carlo simulations \cite{WangMC,Hofstetter},
or a variational approach \cite{GeorgSCHA}. Despite this progress a 
complete theoretical understanding is still missing. 

In this paper we focus on the
frequency dependent conductance of a single metallic tunnel junction
coupled to an electromagnetic environment of arbitrary impedance,
and use a path integral
formulation to analytically calculate the leading order
quantum corrections. 
The semiclassical evaluation of the path integral is justified at high
temperatures and/or for large tunneling conductance and yields the
frequency dependent linear conductance. So far experimental 
data \cite{JoyezSJ} are only available in the zero frequency 
limit where good agreement is found.

\begin{figure}
\begin{center}
\leavevmode
\epsfxsize=0.3 \textwidth
\epsfbox{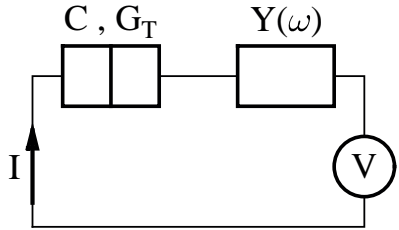}
\end{center}
\caption{Circuit diagram of a tunnel junction in series with an admittance.}
\label{fig:fig1}
\end{figure}

The conductance of a tunnel junction is related
to current fluctuations depending on the whole circuit. Via network
transformations it is always
possible to transform the linear electromagnetic environment into 
an admittance
$Y(\omega)$ in series with a tunnel junction with
tunneling conductance $G_T$ and
capacitance $C$ biased by a voltage source $V$\/, {\it cf}.\ Fig
\ref{fig:fig1}.

\section{Model and linear conductance}
Our interest is in the zero bias differential conductance $G=\partial
I/ \partial V |_{V=0}$ where $I$ is the current in the leads
and $V$ the applied voltage. Employing the Kubo formula
\begin{equation}
G(\omega) =  \frac{1}{i \hbar \omega }  
\lim_{i \nu_n \rightarrow \omega  + i \delta} 
\int^{\hbar \beta}_0 \! d \tau\, e^{i \nu_n \tau} 
\langle I( \tau) I(0)\rangle, 
\label{conduct}
\end{equation}
where the $\nu_n=2 \pi n / \hbar \beta $ are Matsubara frequencies, the
conductance is related to the imaginary time current autocorrelation
function which is most conveniently calculated as a variational
derivative of a generating functional $Z[\xi]$ depending on an auxiliary
field $\xi(\tau)$. Introducing a phase
variable $\varphi$, which is canonically conjugate to the charge
transferred through the junction,
the generating functional can be written as a path integral
\begin{equation}
Z[\xi]  =  \int D[\varphi] \exp  
\left\{- \frac{1}{\hbar} \, S [\varphi, \xi] \right\} \label{pathint}
\end{equation}
with the effective Euclidean action 
\begin{equation}
 S[\varphi,\xi]  =  S_C[\varphi] +S_T[\varphi]  
+ S_Y[\varphi, \xi] \, . \label{action}
\end{equation}
Here
\[
S_C[\varphi] = \int^{\hbar \beta}_0  d \tau \, 
\frac{\hbar^2C}{2e^2} \, \dot{\varphi}^2 
\]
describes Coulomb charging and 
\begin{equation}
S_T[\varphi] = 2 \int^{\hbar \beta}_0 \! d \tau  
\int^{\hbar \beta}_0 \! d \tau^\prime \, \alpha(\tau 
- \tau^\prime) \sin^2 \! \left[\frac{\varphi(\tau)-
\varphi(\tau^\prime)}{2} \right] \label{tunnel}
\end{equation}
quasi-particle tunneling across the junction. 
The kernel $\alpha(\tau)$ is determined by the tunneling 
conductance $G_T$ and may be written as 
\[
\alpha(\tau)  = \frac{1}{\hbar \beta} \sum^{+ 
\infty}_{n=- \infty}  \widetilde{\alpha}(\nu_n) 
\, e^{-i \nu_n \tau}
\]
where the Fourier coefficients are given by
\begin{equation}
 \widetilde{\alpha}(\nu_n) 
  =  
  - \frac{\hbar}{4 \pi} \, 
  \frac{G_T}{G_K} \left| \nu_n \right|  \, .
\label{alphatilde}
\end{equation}
The last term in the action (\ref{action}) describes the effective
environment and includes the auxiliary field $\xi(\tau)$
\begin{equation}
 S_Y[\varphi, \xi] 
 =
 \frac{1}{2} \int^{\hbar 
 \beta}_0  d \tau  \int^{\hbar \beta}_0  d \tau^\prime 
 k(\tau - \tau^\prime) 
 \left[\varphi(\tau) + \frac{e}{\hbar}\xi(\tau) -
 \varphi(\tau^\prime ) - 
 \frac{e}{\hbar} \xi(\tau^\prime)\right]^2,
\label{resistor}
\end{equation}
where the kernel $k(\tau)$ can also be written as a Fourier series 
with coefficients
\begin{equation}
 \widetilde{k}(\nu_n)
  =
  -\frac{\hbar}{4\pi} \frac{\widehat{Y}(|\nu_n|)}{G_K} |\nu_n|  .
\end{equation} 
Here $\widehat{Y}(s)$ denotes the Laplace transform of the
environmental response 
function, {\it cf.} Ref.\ \cite{GrabertREP}. Due to
causality, for $\mbox{Re}(s)>0$, one may write  
$\widehat{Y}(s)=Y(is)$ where
$Y(\omega)$ is the frequency dependent admittance of the
environment. 

We now perform the functional
derivatives explicitly and get for the correlator \cite{GeorgSJ}
\begin{equation}
 \langle I(\tau)I(0)\rangle 
 =
 \frac{1}{Z}  \int 
 D[\varphi] 
 \exp \left\{ - \frac{1}{\hbar} S[\varphi,0] \right\}
 \left\{ 2 \frac{e^2}{\hbar} k(\tau) + I [\varphi, \tau] 
 I [\varphi, 0] 
 \right\},
\label{phicorr} 
\end{equation}
where $Z=Z[0]$ denotes the partition function and 
the current functional $I[\varphi, \tau]$ is given by 
\begin{equation}
I [\varphi, \tau] = \frac{2e}{\hbar} 
\int^{\hbar \beta}_0 d \tau^\prime \,  k(\tau - \tau^\prime) 
\varphi(\tau^\prime) \, .
\label{currfunc}
\end{equation} 
The first term in $(\ref{phicorr})$ is independent of the phase 
variable $\varphi$ and
can be handled exactly, whereas the second term cannot be
evaluated without further approximations. Thus, we split the
conductance in two parts $G=G_1 + G_2$. 
The first part stemming from the second order functional derivative 
of the action reads
\begin{equation}
 G_1(\omega) 
 =
 \frac{1}{i \hbar \omega} \frac{2 e^2}{\hbar} 
  \widetilde{k}(-i \omega+\delta)
 =
 Y(\omega),  
\end{equation}
where the unique continuation \cite{Baym} of $\widetilde{k}(\omega)$ 
is obtained by
defining the analytic continuation of the absolute value as
\begin{equation}
 |z| = 
  \left\{
   \begin{array}{c@{\qquad}c}
    z  & \mbox{Re}(z)>0   \\
    -z  &  \mbox{Im}(z)<0 
   \end{array}
  \right._.
\end{equation}
The second part $G_2$ of the conductance
 results from a product of two first order  
derivatives of the action, and we write
\begin{equation}
 G_2(\omega)
 =
 \frac{1}{i \hbar \omega } C_2(-i\omega + \delta)
\end{equation}
with  
\begin{equation}
 C_2(\nu_n)
 =
 \frac{1}{Z}\int D[\varphi] \exp
 \left\{
  -\frac{1}{\hbar}S[\varphi,0]
 \right\}
 F[\varphi,\nu_n],
\label{eq:C2formal}
\end{equation}
where the explicit form of the current functional 
$(\ref{currfunc})$ written
in terms of the Fourier transform of the phase yields
\begin{equation}
 F[\varphi,\nu_n]
 =
 \frac{4 e^2 \beta}{\hbar} \widetilde{k}(\nu_n) \widetilde{\varphi}(\nu_n)
 \sum_{m=-\infty}^{+\infty} \widetilde{k}(\nu_m)
 \widetilde{\varphi}(\nu_m).
\end{equation}
This is a formally exact representation of the linear conductance.

\section{Semiclassical limit}
\begin{figure}
\begin{center}
\leavevmode
\epsfxsize=0.3 \textwidth
\epsfbox{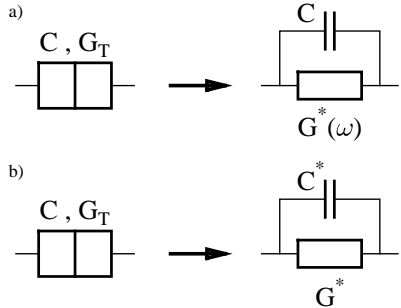}
\end{center}
\caption{Effective circuit diagrams for a tunnel junction in 
the semiclassical limit a) for arbitrary frequency and b) in the 
low frequency limit.}
\label{fig:fig2}
\end{figure}

To proceed we 
expand the effective action around the classical path. Following the
lines of Ref.\ \cite{GeorgSJ} we change to Fourier space
diagonalizing the second order variational action. The eigenvalues 
are given by 
\begin{equation}
 \lambda(\nu_n)
 =
 \frac{\hbar^2 \beta}{e^2}
 |\nu_n| \left[ \widehat{G}_0(\nu_n) + \widehat{Y}(|\nu_n|) \right],
\label{lambdatilde}
\end{equation}
where
\begin{equation}
 \widehat{G}_0(\nu_n) = |\nu_n| C  + G_T 
\end{equation}
describes the junction as a capacitance in parallel with an Ohmic
resistor characterized by the classical tunneling resistance. 
Including the fourth order variational derivative of the action we 
get from
$(\ref{eq:C2formal})$ the expression 
\begin{equation}
 C_2(\nu_n) 
 =
 \frac{4e^2\beta}{\hbar}\,
 \frac{\widetilde k(\nu_n)^2}{\lambda(\nu_n)}
 \Bigg[
   1 + \frac{2\beta}{\lambda(\nu_n)}
   \sum_{{m=-\infty}\atop {m\ne 0}}^{\infty}
   \frac{\widetilde\alpha(\nu_{n+m})-\widetilde\alpha(\nu_n)
          -\widetilde\alpha(\nu_m)}{\lambda(\nu_m)}
 \Bigg]  .
 \label{eq:C_2}
\end{equation}
The convergence of this expansion depends crucially on the eigenvalues
$(\ref{lambdatilde})$. To estimate the validity, we consider the
smallest eigenvalue in more appropriate units
\begin{equation}
 \lambda(\nu_1) 
 =
 \left[
  \frac{2 \pi^2}{\beta E_C} + \frac{G_T+\widehat{Y}(\nu_1)}{G_K}
 \right] 
\end{equation}
where $\beta$ denotes the inverse temperature
and $E_C=e^2/2C$ the charging energy.
This eigenvalue has to be large compared to $1$, and we see that the 
expansion is useful for large conductance
$G_T + \widehat{Y}(\nu_1) \gg G_K$ and/or high temperatures
$\beta E_C \ll 2 \pi^2$.
We now perform the limit $i\nu_n \rightarrow \omega
+i\delta$. Consider the analytically continued eigenvalue 
\begin{equation}
 \lambda(-i \omega)
 =
 -i \omega \frac{\hbar^2 \beta}{e^2}
 \left[
 G_0(\omega) + Y(\omega)
 \right],
\label{eq:lambdacont}
\end{equation}
where
\begin{equation}
 G_0(\omega)
 = 
 G_T -i\omega C
\end{equation}
is the continuation of the Laplace transform 
$\widehat{G}_0(-i\omega + \delta )$.
The relative minus sign of the capacitive term is due to the usual
definition of the Fourier transform in quantum mechanics, the 
electro-technical convention is obtained by replacing
$\omega \rightarrow -\omega$. 
For small frequencies
the continued eigenvalue 
$(\ref{eq:lambdacont})$ is
no longer large compared to $1$
and we are faced with a problem of order reduction. This 
is handled systematically in Ref.\ \cite{GeorgSJ} showing 
that inclusion of the fourth order variation is sufficient up 
to first order in $\beta E_C$ and
$G_K/(G_T +  \widehat{Y}(\nu_1))$, respectively. In the
reminder we omit the order symbol, but the
meaning of the equations is always meant in a limiting sense up to 
this order. After performing the
continuation we get for the conductance
\begin{equation}
 G(\omega)
 =
 \frac{G_{\rm eff}(\omega) Y(\omega)}
      {G_{\rm eff}(\omega) + Y(\omega)}
\label{eq:Gallg}
\end{equation}
with an effective linear conductance of the junction
\begin{equation}
 G_{\rm eff}(\omega)
 =
 G_T
 \left[
   1 - {\cal U}(\omega) 
 \right]
   -i \omega C  .
\end{equation}
This describes a linear element 
$G^*(\omega)=G_T[1-{\cal U}(\omega)]$, depending on the whole circuit, in 
parallel with the junction capacitance $C$ as 
depicted in Fig.\ \ref{fig:fig2}a. Here 
\begin{equation}
 {\cal U}(\omega)
 =
 \frac{2}{i \omega} \sum_{m=1}^\infty \nu_m 
 \left[
  \frac{1}{\lambda(\nu_m - i \omega)} -\frac{1}{\lambda(\nu_m)}
 \right]
\label{eq:qmsupp}
\end{equation}
describes the suppression of the conductance due to discrete charge 
transfer. The general form $(\ref{eq:Gallg})$ is valid only 
to first order in $\beta E_C$ and 
$G_K/(G_T + \widehat{Y}(\nu_1))$, respectively. 
A systematic treatment of higher order contributions does
not allow for a description of the tunnel junction in terms of 
an effective linear element depending on the whole circuit. 
However, a partial resummation of higher order terms performed by 
a self consistent harmonic approximation \cite{GeorgSCHA}
leads again to the form $(\ref{eq:Gallg})$.

\section{Discussion} 
\begin{figure}
\begin{center}
\leavevmode
\epsfxsize=0.45 \textwidth
\epsfbox{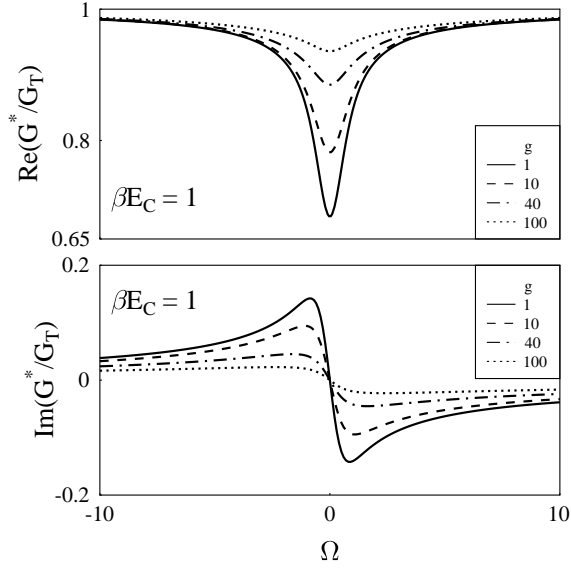}
\end{center}
\caption{Real and imaginary parts of $G^*(\omega)/G_T$ in the ohmic damping
case for $\beta E_C=1$ and dimensionless conductance
$g=1,10,40$ and $100$ in dependence on the dimensionless frequency
$\Omega = \hbar \omega /2 \pi E_C$.}
\label{fig:fig3}
\end{figure}

For further discussion of the conductance and a comparison with 
experimental data we now
restrict ourselves to ohmic dissipation $Y(\omega)=Y$. 
We then get for the effective linear element 
\begin{equation}
 \frac{G^*(\omega)}{G_T}
 =
 1-
 \Big[
  \frac{\psi(1+u+\tilde{\omega})
        -\psi(1+\tilde{\omega})}{u}
  +
  \frac{\psi(1+u+\tilde{\omega})
        -\psi(1+u)}{\tilde{\omega}}
 \Big] \frac{\beta E_C}{\pi^2},
\label{eq:ohmomega}
\end{equation}
where $\psi$ is the logarithmic derivative of the gamma function and
\begin{equation}
 u
 =
 g \frac{\beta E_C}{2\pi^2} 
 , \qquad
 \tilde{\omega}=\frac{\hbar \beta}{2 \pi i} \omega 
\end{equation}
are auxiliary quantities.
We also have introduced the dimensionless parallel conductance 
$g=(G_T + Y)/G_K$. The quantum corrections depend only on this
combination of conductances.
The real and imaginary parts of $G^*(\omega)/G_T$ are depicted in
Fig.\ \ref{fig:fig3} for $\beta E_C=1$ and various values of $g$.

Due to the logarithmic behavior of the psi function for 
large arguments, the quantum corrections disappear nonanalytically for
large $\omega$. 
For small frequencies the effective element behaves like a
renormalized admittance with an additional capacitance in parallel and we
may define a renormalized capacitance $C^*$ and a renormalized conductance
$G^*=G^*(\omega=0)$, {\it cf}.\ Fig.\ \ref{fig:fig2}b. 
For the renormalized conductance we get
\begin{equation}
 \frac{G^*}{G_T}
 = 
 1-
 \left[
  \frac{\gamma+\psi(1+u)}{u} + \psi'(1+u)
 \right] \frac{\beta E_C}{\pi^2}
\label{eq:ohm0}
\end{equation}
which coincides with our previous result
\cite{GeorgSJ}. $G^*$ exhibits a nonanalytic behavior
in the limit of vanishing environmental resistance. 

\begin{figure}
\begin{center}
\leavevmode
\epsfxsize=0.55 \textwidth
\epsfbox{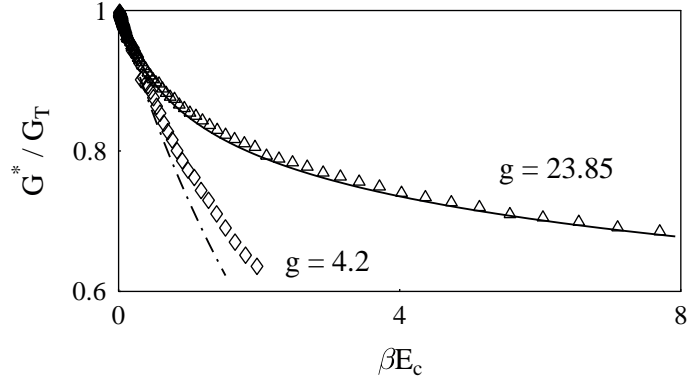}
\end{center}
\caption{The linear conductance versus the dimensionless 
temperature for two dimensionless parallel
conductance $g=4.2$ and $23.85$ compared with experimental data 
(symbols) by Joyez {\it et al}.\ \protect\cite{JoyezSJ}.}
\label{fig:fig4}
\end{figure}

We compare our prediction $(\ref{eq:ohm0})$ with recent 
experimental
data by Joyez {\it et al}.\ \cite{JoyezSJ} for 
dimensionless conductance
$g=4.2$ and $23.85$. 
Fig.\ \ref{fig:fig4} shows
that in the limit of large conductance we are able to cover the whole
temperature range  explored experimentally with no adjustable
parameter, whereas for 
moderate conductance only the high
temperature part is covered by the semiclassical theory.

The renormalized capacitance $C^*$ incorporates the linear part in
$\omega$ of the imaginary part of
$G^*(\omega)$ and the geometrical capacitance $C$, and we get 
\begin{equation}
 \frac{C^*}{C}
 =
 1+ \frac{G_T}{G_K}
 \left[
  \frac{\frac{\pi^2}{6} -\psi'(1+u)}{2 \pi^2 u}
  -\frac{\psi''(1+u)}{4 \pi^2}
 \right]
 \frac{(\beta E_C)^2}{\pi^2}.
\label{eq:cap0}
\end{equation} 
The correction shows a quadratic dependence on 
$\beta E_C$. The renormalization 
is suppressed at high temperatures and also vanishes linearly for
large conductance.

In summary, we have derived an analytical expression for the frequency
dependent linear conductance of a tunnel junction in the semiclassical
limit. We have shown
that this limit covers not only high temperatures but also large 
conductance and agrees with experimental findings.

The authors would like to thank Michel Devo\-ret, Daniel Esteve, and 
Philippe Joyez for valuable discussions.
One of us (GG) acknowledges the hospitality of the CEA-Saclay 
during an extended stay.
Financial support was provided by the Deutsche
Forschungsgemeinschaft (DFG) and the Deutscher Akademischer 
Austauschdienst (DAAD).

%\end{multicols}
\end{document}